\documentclass[twocolumn]{aastex631}

\usepackage{acronym}
\usepackage{courier}
\usepackage{multirow}
\usepackage{amsmath}
\usepackage{xspace}

\acrodef{SED}{spectral energy distribution}
\acrodef{DDF}{deep drilling field}
\acrodef{LSST}{Legacy Survey of Space and Time}
\acrodef{AGN}{active galactic nucleus}
\acrodef{GALEX}{Galaxy Evolution Explorer}
\acrodef{VOICE}{VST Optical Imaging of the CDF-S and ELAIS-S1 Fields}
\acrodef{HSC}{Hyper Suprime-Cam}
\acrodef{VIDEO}{VISTA Deep Extragalactic Observations}
\acrodef{DeepDrill}{Spitzer Survey of Deep-Drilling Fields}
\acrodef{SWIRE}{Spitzer Wide-area Infrared Extragalactic survey}
\acrodef{DES}{Dark Energy Survey}
\acrodef{SFR}{star formation rate}
\acrodef{sSFR}{specific \ac{SFR}}
\acrodef{OACAPI}{Open Astronomy Catalogs API}
\acrodef{SNe}{supernovae}
\acrodef{SN}{supernova}
\acrodef{MLP}{multilayer perceptron}
\acrodef{NN}{neural network}
\acrodef{ReLU}{rectified linear unit}
\acrodef{MSE}{mean squared error}
\acrodef{RMSE}{root mean squared error}
\acrodef{M_s}[$M_*$]{stellar mass}
\acrodef{RF}{random forest}
\acrodef{SMOTE}{Synthetic Minority Oversampling Technique}
\acrodef{KNN}{k-nearest neighbors}
\acrodef{SPLASH}[\texttt{SPLASH}]{Supernova classification Pipeline Leveraging Attributes of Supernova Hosts}
\acrodef{OVR}{one-versus-rest}
\acrodef{CCSN}{core collapse supernova}
\acrodef{CCSNe}{core collapse supernovae}
\acrodef{SLSN}{superluminous supernova}
\acrodef{SLSNe}{superluminous supernovae}
\acrodef{DR2}{Data Release 2}
\acrodef{MAST}{Mikulski Archive for Space Telescopes}
\acrodef{BTS}{Bright Transient Survey}
\acrodef{YSE}[YSE DR1]{Young Supernova Experiment Data Release 1}
\acrodef{ZTF}{Zwicky Transient Facility}
\acrodef{Pan-STARRS}{Panoramic Survey Telescope and Rapid Response System}
\acrodef{GLADE}{Galaxy List for the Advanced Detector Era}
\acrodef{DECaLS}{Dark Energy Camera Legacy Survey}
\acrodef{4MOST}{4-metre Multi-Object Spectroscopic Telescope}
\acrodef{TiDES}{Time Domain Extragalactic Survey}

\newcommand{\photozs}{photo-$z$s\xspace}

\newcommand{\specz}{spec-$z$\xspace}
\newcommand{\speczs}{spec-$z$s\xspace}
\newcommand{\logM}{$\log(M_*)$\xspace}
\newcommand{\logSFR}{$\log(\rm{SFR})$\xspace}
\newcommand{\z}{$z$\xspace}
\newcommand{\IAIFI}{The NSF AI Institute for Artificial Intelligence and Fundamental Interactions}
\newcommand{\MIT}{Department of Physics and Kavli Institute for Astrophysics and Space Research, Massachusetts Institute of Technology, 77 Massachusetts Avenue, Cambridge, MA 02139, USA}
\newcommand{\CfA}{Center for Astrophysics $|$ Harvard \& Smithsonian, Cambridge, MA 02138, USA}
\newcommand{\pstarr}{\ac{Pan-STARRS}\xspace}

\newcommand{\calcfactor}[1]{%
  \dimexpr#1\textwidth-2\tabcolsep-1.5\arrayrulewidth\relax
}
\newcolumntype{P}[1]{p{\calcfactor{#1}}}

\begin{document}

\title{SPLASH: A Rapid Host-Based Supernova Classifier for Wide-Field Time-Domain Surveys}

\author[0009-0005-9830-9966]{Adam Boesky}
\affiliation{\CfA}

\author[0000-0002-5814-4061]{V.~Ashley~Villar}
\affiliation{\CfA}
\affiliation{\IAIFI}

\author[0000-0003-4906-8447]{Alexander~Gagliano}
\affiliation{\IAIFI}
\affiliation{\CfA}
\affiliation{\MIT}

\author[0000-0002-9454-1742]{Brian~Hsu}
\affiliation{Steward Observatory, University of Arizona, 933 North Cherry Avenue, Tucson, AZ 85721-0065, USA}

\begin{abstract}

The upcoming Legacy Survey of Space and Time (LSST) conducted by the Vera C. Rubin Observatory will detect millions of supernovae (SNe) and generate millions of nightly alerts, far outpacing available spectroscopic resources.
Rapid, scalable photometric classification methods are therefore essential for identifying young SNe for follow-up and enabling large-scale population studies.
We present SPLASH, a host-based classification pipeline that infers supernova classes using only host galaxy photometry.
SPLASH first associates SNe with their hosts (yielding a redshift estimate), then infers host galaxy stellar mass and star formation rate using deep learning, and finally classifies SNe using a random forest trained on these inferred properties, along with host-SN angular separation and redshift.
SPLASH achieves a binary (Type Ia vs. core-collapse) classification accuracy of 76\% and an F1-score of 69\%, comparable to other state-of-the-art methods.
By selecting only the most confident predictions, SPLASH can return highly pure subsets of all major SN types, making it well-suited for targeted follow-up.
Its efficient design allows classification of ~500 SNe per second, making it ideal for next-generation surveys.
Moreover, its intermediate inference step enables selection of transients by host environment, providing a tool not only for classification but also for probing the demographics of stellar death.

\end{abstract}


\section{Introduction} \label{sec:intro}

The explosive deaths of stars, called \Ac{SNe}, are fundamental to the composition, structure, dynamics, and evolution of the Universe.
The Vera C.~Rubin Observatory is set to begin the \ac{LSST} in $2025$, a decade-long survey that is expected to photometrically discover over $1$ million \ac{SNe} each year \citep{LSST_2009}.
The \ac{SNe} that Rubin detects will be buried within an unprecedented $10$ million \textit{nightly} transient alerts.
While Rubin's extraordinary rate of \ac{SN} detection will revolutionize the volume and diversity of the known transient catalog, the \ac{SN} detections that it produces must be swiftly categorized and sorted to select the most interesting candidates for observational follow up and to study transient subpopulations while they remain bright.
Transient classification will be an essential first step in sifting through the terabytes of data collected each night.

\ac{SN} classification is a rapidly evolving field whose methodology has seen considerable improvement in recent years.
Historically, \ac{SNe} were classified spectroscopically, which remains the gold standard \citep[e.g.,][]{Filippenko_1997}.
However, spectroscopic observations are resource-intensive.
This has motivated a shift toward photometric classification methods, often for identifying promising candidates for spectroscopic follow-up.
The advent of \ac{LSST} will produce so many \ac{SN} alerts that comprehensive spectroscopic follow-up will be untenable.
Currently, only $\sim 1/10$ detected transient phenomena are followed spectroscopically \citep{kulkarni_2020}.
Assuming that observational resources remain the same, this statistic will plummet to $\sim 1/500$ with the onset of \ac{LSST}, although upcoming multi-fiber spectroscopic surveys such as \ac{4MOST} \ac{TiDES} will increase our expected follow-up rate by factors of $\sim2-3$ \citep{frohmaier2025tides}.
To maximize scientific discovery in the coming years, it is therefore essential to develop rapid, efficient, and accurate methods for classifying photometric transient alerts.

The early stages of \ac{SNe} are particularly important for constraining the physics of their progenitor events and understanding the demographics of the supernova population. 
Many fundamental questions about early-time \ac{SN} behavior remain unresolved, including the physical mechanisms driving their emission and how these mechanisms relate to progenitor environments. 
For example, young Type II \ac{CCSNe} like \ac{SN}\ 2023ixf \citep{Bostroem_2023} show signs of interaction with dense circumstellar material ejected in the months to years prior to explosion.
Meanwhile, the cause of the early luminosity peaks seen in \ac{SLSNe} remains poorly understood, with no consensus model to explain them \citep{Zhu_2023}.

Early photometric classification can trigger timely spectroscopic follow-up, capturing \ac{SNe} at a stage when unique physical insights like mass loss kinematics in Type II events or signatures of the central engine in \ac{SLSNe} are still accessible.
However, classification at these early stages is especially challenging due to the limited information available from just a few days of photometric data \citep{gagliano_2023}.
Luckily, \ac{SNe} come from a variety of galactic environments with distinct underlying stellar populations and evolutionary trajectories leading up to explosion \citep{Leaman_2011, Kelly_2012, Hakobyan_2012, Childress_2013, Kisley_2023, Villar_2025}.
We may therefore be able to leverage known galaxy-\ac{SN} correlations to aid in early, cheap, and accurate classification.

In this paper, we present a host-based photometric machine learning pipeline called \ac{SPLASH}.
\ac{SPLASH} uses multi-band photometry to infer the physical properties of host galaxies, and then it classifies given \ac{SNe} based on the predicted properties of their hosts.
\ac{SPLASH} is optimized for early-time classification because it uses only host information, meaning that it can infer a \ac{SN}'s class from the instant it appears in the sky.

While \ac{SN} classification using multi-band host galaxy photometry \textit{or} derived properties has been attempted before (e.g. \citet{Kisley_2023} and \citet{astro_ghost}) our pipeline is a novel unification of these methods, going from photometry to inferred host properties, and then to \ac{SN} class.
Inferring properties as an intermediary step will enable scientists to use \ac{SPLASH} to select for specific host demographics such as host galaxies with high stellar mass or at low redshifts.
Philosophically, inferring host properties also makes more sense than alternatives that classify using only photometry---\ac{SPLASH}'s structure reflects the notion that there is a real, physical mapping between photometry and properties, and a coupling between galaxy attributes and the \ac{SNe} within them.

The structure of this paper is as follows:
Section~\ref{subsec:datasets} describes the datasets that we use and our procedure for compiling them, Section~\ref{subsec:host_prop_NN} and Section~\ref{subsec:random_forest} describe the methodology of \ac{SPLASH}, and Section~\ref{sec:pipeline_performance} evaluates its performance using the metrics described in Section~\ref{subsec:evaluation_metrics}.
The accuracy of \ac{SPLASH}'s galaxy property prediction is described in Section~\ref{subsec:host_prop_NN_performance}, and we assess the performance of our random forest classification in Section~\ref{subsec:rf_performance}.
Finally, we discuss the use cases and performance of \ac{SPLASH} relative to other cutting-edge photometric classification models in Section~\ref{sec:discussion}. 

\begin{figure*}
  \includegraphics[width=\linewidth]{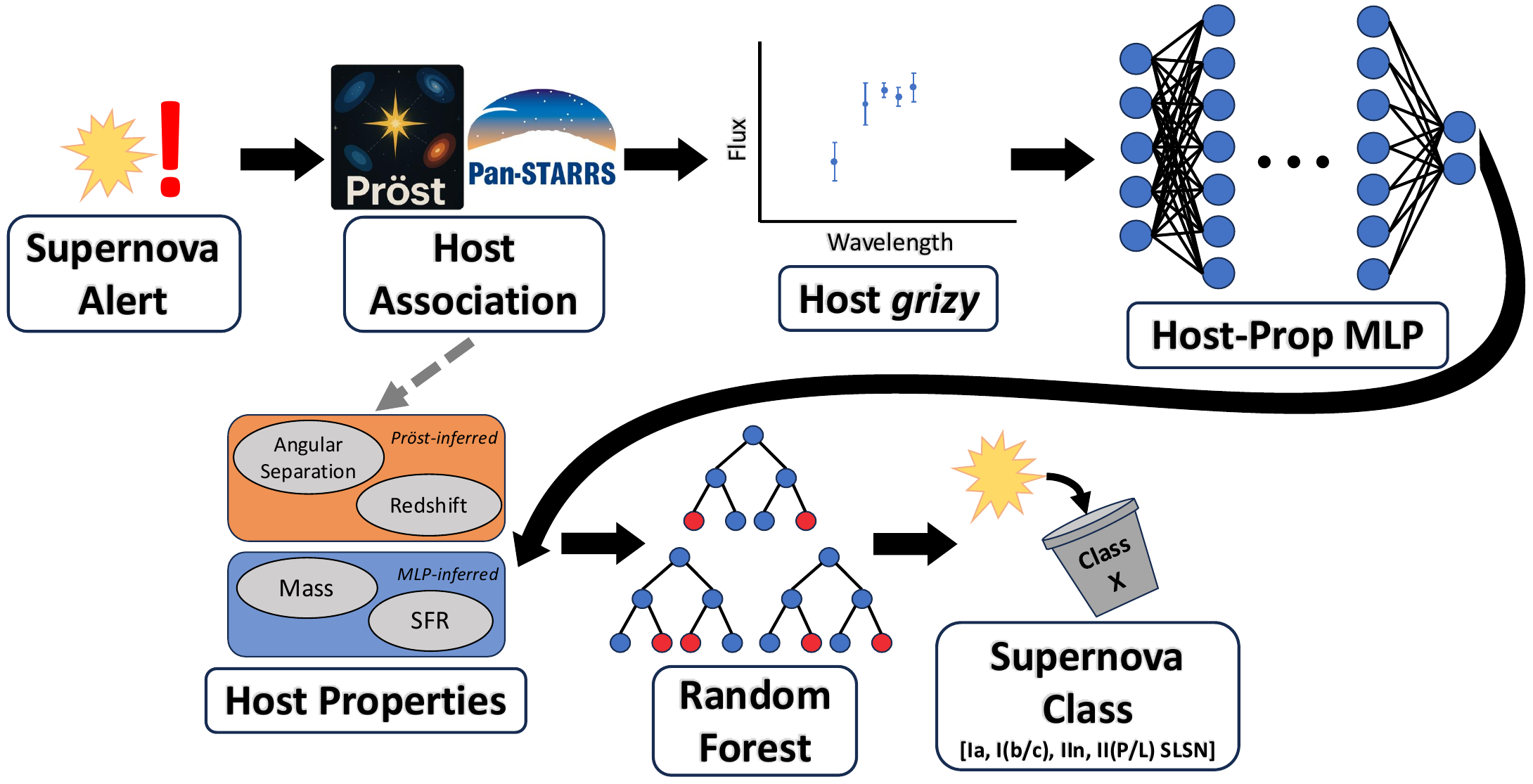}
  \caption{The \ac{SPLASH} pipeline architecture. When a supernova alert occurs, it is associated with a galaxy in the \pstarr catalog using the \texttt{Pröst} probabilistic host galaxy association software.
  If a redshift is not provided in the catalog, \texttt{Pröst} will provide an estimate of the host's redshift.
  The host galaxy's \textit{grizy} photometry is then fed into a multilayer perceptron that infers its stellar mass and star formation rate.
  These inferred properties, the redshift, and the host-SN angular separation are then passed into a random forest which classifies the SN. The classes of the supernovae that we infer are Type Ia, I(b/c), IIn, II(P/L), and SLSNe.}
  \label{fig:methods_figure}
\end{figure*}

\section{Methods} \label{sec:methods}

\ac{SPLASH} uses photometric measurements of host galaxies to classify \ac{SNe}.
The structure of this methods section mirrors that of our pipeline, which is displayed in Figure~\ref{fig:methods_figure}: Section~\ref{subsec:datasets} describes the datasets we use for training and testing, Section~\ref{subsec:host_association} summarizes how we associate supernovae with their host galaxies, Section~\ref{subsec:host_prop_NN} describes how we apply deep learning to infer host properties, and Section~\ref{subsec:random_forest} details how we classify \ac{SNe} using their inferred host properties.
Finally, in Section~\ref{subsec:evaluation_metrics} we define the metrics that we use to evaluate the performance of \ac{SPLASH}.

\subsection{Datasets} \label{subsec:datasets}

\subsubsection{Host Property Data} \label{subsubsec:gal_catalog}

We use the dataset from \citet{Zou_2022} to train the host property inference component of \ac{SPLASH}.
The dataset provided by \citet{Zou_2022} is ideal for our use case because it provides optical photometric measurements tagged with physical properties for galaxies within several of \ac{LSST}'s \acp{DDF}, making it particularly relevant for the era of the Vera C. Rubin Observatory.
\citet{Zou_2022} provides a catalog of multi-wavelength observations, \ac{SED} fits, and derived physical properties for $2,873,803$ galaxies.
These galaxies are from W-CDF-S ($4.9 \ \rm{deg}^2$), ELAIS-S1 ($3.4 \ \rm{deg}^2$), and XMM-LSS ($4.9 \ \rm{deg}^2$), three \acp{DDF} of Rubin's \ac{LSST}.
The archival photometric observations are collected from the \ac{GALEX} \citep{Martin_2005}, \ac{VOICE} \citep{Vaccari_2016}, \ac{HSC} \citep{Ni_2019}, \ac{VIDEO} \citep{Jarvis_2013}, \ac{DeepDrill} \citep{Lacy_2021}, \ac{SWIRE} \citep{Lonsdale_2003, Surace_2012}, and \ac{DES} \citep{Abbott_2021} catalogs.
To make our model applicable to forthcoming \ac{LSST} data, we use the $g$, $r$, $i$, $z$, and $y$ filters.

By fitting source \acp{SED} from the X-ray to the far infrared with \texttt{CIGALE} \citep{Boquien_2019}, \cite{Zou_2022} compiles a catalog of derived galaxy properties.
\texttt{CIGALE} predicts galactic \ac{M_s} and \ac{SFR}.
Although \texttt{CIGALE} can estimate redshift, \citet{Zou_2022} does not use it for validation due to concerns about parameter degeneracies and limited validation of its photo-\textit{z} performance.
Photometric redshifts (\photozs) and spectroscopic redshifts (which are used when available) are therefore compiled from \citet{Ni_2021} and \citet{Zou_2021}.
The \photozs are estimated using \texttt{EAZY} \citep{EAZY}.
\citet{Zou_2021} also uses specially-tailored models for fitting the \acp{SED} of galaxies that are promising candidates for containing an \ac{AGN}.
Each galaxy is fit both with and without the \ac{AGN}-specific model, and the probability of an \ac{AGN} is provided.
We use \ac{M_s} and \ac{SFR} from the most likely of these two fits for each galaxy.

Before training our model, we preprocess the data from \citet{Zou_2022}.
First, we filter out all galaxies with photometric or spectroscopic redshift estimates beyond $z=1$.
\ac{SN} detections beyond this limit will be rare even with \ac{LSST} (with the exception of SLSNe), and, more importantly, the properties of high-$z$ galaxies are likely poorly constrained by \texttt{CIGALE}.
Second, we match the bands from each survey based on their approximate median effective wavelength.
For example, the \textit{r}-bands for ELAIS-S1, W-CDF-S, and XMM-LSS come from \ac{DES}, \ac{VOICE}, and \ac{HSC}, respectively, but we use them interchangeably because their bandpasses are similar enough for our purposes.
Third, if there is more than one catalog with observations in a given band, we take the values from the catalog with fewer missing observations across all galaxies for that band.
Fourth, and finally, we use \ac{KNN} imputation, a method that replaces missing data points with the mean of the closest $k$ values.
We set $k=5$ to account for missing photometric measurements.
The fraction of values that needs to be imputed varies between filters, but is always less than $10\%$ and typically less than $5\%$.

\subsubsection{\pstarr Host Data} \label{subsubsec:PSTAR_catalog}

We use \pstarr \ac{DR2}, which we refer to as simply ``\pstarr" throughout this paper, as a host galaxy catalog to associate \ac{SNe}. \ac{DR2} is the second data release from the $3\pi$ Survey conducted by the \pstarr wide-field astronomical imaging system.
\ac{DR2} substantially overlaps both our training set and ongoing surveys and has similar photometric bands to \ac{LSST}.
The catalog consists of stacked images, mean attribute catalogs, and static sky catalogs from the $3\pi$ Survey in \textit{grizy} from the $3/4$ of sky north of declination $-30^\circ$ \citep{panstarrs_surveys}.
\ac{DR2} consists of data collected from $2010$-$2014$, and includes more than $10$ billion objects.
To access \pstarr data, we use the \ac{MAST} API\footnote{\url{https://catalogs.mast.stsci.edu/docs/index.html}}.
The \pstarr \ac{DR2} data used in this paper can be found in MAST: \dataset[10.17909/s0zg-jx37]{http://dx.doi.org/10.17909/s0zg-jx37}.
See \citet{panstarrs_surveys} for a summary of the \pstarr survey design and \citet{panstarrs_data} for a description of the \pstarr data storage and API.

\subsubsection{The Supernova Dataset} \label{subsubsec:sne_catalog}

We use the \ac{OACAPI} \citep{OAC} to retrieve a compiled dataset of \ac{SNe} from heterogeneous sources including wide-field synoptic surveys, individual publications, and other transient catalogs like the Transient Name Server\footnote{\url{wis-tns.org}} and Gaia Photometric Science Alerts\footnote{\url{gsaweb.ast.cam.ac.uk/alerts/home}}.
\ac{OACAPI} aggregates \ac{SN} coordinates, photometry, spectroscopic classifications, host galaxy associations, redshifts, and other metadata for each event.
Our dataset consists of $82,605$ SN candidates, $17,319$ of which have a spectroscopic classification from the literature.
There are $965$ SNe with host galaxies in the \acp{DDF} of interest, $586$ of which are classified and $\sim 75\%$ of which are Type Ia.
Henceforth, we shall refer to this set of classified supernovae with hosts as ``the supernova dataset". 

In Figure~\ref{fig:pairplot}(b), we show the distributions of inferred host galaxy properties for each class of \ac{SN}.
Differences in the host property distributions between classes in Figure~\ref{fig:pairplot}(b) confirm the correlation between \ac{SN} classes and the properties of their hosts.
For instance, Type Ia are observed across all galaxy types while \ac{CCSNe} only occur in galaxies with ongoing or recent star formation \citep{Li_2011, Hakobyan_2012, Childress_2013, Schulze_2021, Qin_2024, Villar_2025}.
Within \ac{CCSNe}, Type Ib/c \ac{SNe} prefer high-mass and high-metallicity galaxies compared to Type II \citep{Kelly_2012, Schulze_2021, Qin_2024}.
Rare \ac{SNe} typically come from exotic host galaxies; \ac{SLSNe}, in particular, are found in low-mass, low-metallicity hosts with robust star formation and occur at low offsets due to their high luminosity allowing for distant observation \citep{Kelly_2012, Schulze_2021, Villar_2025}.

Although \ac{SN} classes are correlated with their environments, the overlap in distributions in Figure~\ref{fig:pairplot}(b) demonstrates that they are not strictly linearly differentiable.
Overlap in host feature space makes the task of host-based classification difficult, which we discuss further in Section~\ref{sec:pipeline_performance}.

\subsubsection{BTS and YSE Supernova Data} \label{subsubsec:BTS_YSE_catalog}

\ac{OACAPI} is the primary \ac{SN} dataset used throughout this study, but we use \ac{SNe} from the \ac{BTS} and \ac{YSE} to validate our results.
\ac{BTS} is a public, magnitude-limited ($m < 19$) catalog of transient properties and spectroscopic classes from the \ac{ZTF} \citep{ztf}.
When we downloaded the data on March $1$st, $2024$, \ac{BTS} contained $5343$ spectroscopically classified transients.
\ac{BTS} data is publicly available\footnote{\url{https://sites.astro.caltech.edu/ztf/bts/bts.php}}, and \citet{BTS} describes the methodology and demographics of the catalog.

\ac{YSE} contains photometry, host-galaxy associations, redshifts, classifications, and more information about $1975$ transients.
The catalog is comprised of mostly \ac{SN}e which extend out to $z \approx 0.5$.
The data is publicly available for download\footnote{\url{https://doi.org/10.5281/zenodo.7317476}}, and for more about \ac{YSE} and its methodology see \citet{YSE}.
Throughout this study, we combine the BTS and YSE datasets, and refer to it as ``BTS \& YSE''.

\begin{figure*}
  \centering
  \includegraphics[width=\linewidth]{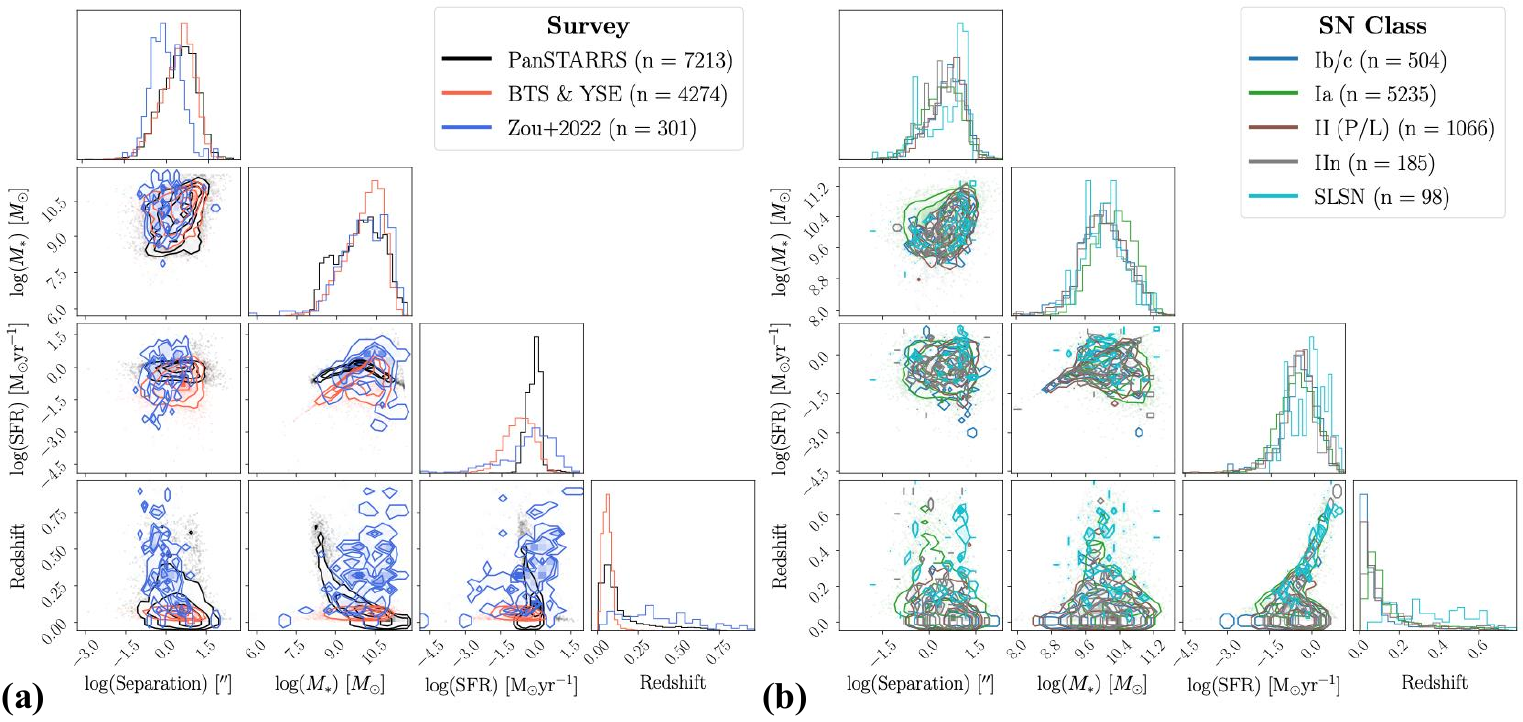}
  \caption{\textbf{(a)} The distributions of SN-host angular separation, stellar mass, star formation rate, and redshift for all hosts from the \pstarr, BTS \& YSE, and \citet{Zou_2022} datasets. Note that the values for \pstarr and BTS \& YSE are inferred by \ac{SPLASH} while the values from \citet{Zou_2022} are derived from their SED fits.
  \textbf{(b)} The distribution of inferred host properties for Type Ia, Ib/c, II, IIn, and SLSNe with hosts in the \pstarr dataset.
  We only include hosts with redshifts $\leq 1$ because \ac{SPLASH} does not classify \ac{SNe} that are farther away.
  All histograms are normalized, and the number of hosts for each category (n) is included in the legends.}
  \label{fig:pairplot}
\end{figure*}

We associate all \ac{SNe} from BTS and YSE with hosts in \pstarr using the probability of chance coincidence method described in Section~\ref{subsubsec:Pcc}.
In Figure~\ref{fig:pairplot}(a), we show the pairwise distributions of host demographics for \pstarr, BTS, YSE, and \citet{Zou_2022}.
The distributions of \pstarr and BTS \& YSE \ac{SNe} are quite close, which is expected as they are compiled from similar sources.
\citet{Zou_2022}, however, contains higher redshift objects with lower separations and a wider spread of \ac{SFR} and \ac{M_s}.
These discrepancies come from the fact that \citet{Zou_2022} has a higher limiting magnitude than the other catalogs.
This allows for the inclusion of deeper galaxies that appear closer to their SNe because of how far they are from the observer, but also galaxies with properties that lead to very dim appearance.

\subsection{Supernova-Host Association}
\label{subsec:host_association}

To associate \ac{SNe} with their host galaxies, \ac{SPLASH} uses the \texttt{Pröst} Python package\footnote{\url{https://github.com/alexandergagliano/Prost}} (Gagliano et al., in prep.).
\texttt{Pröst} associates transients with host galaxies by estimating the posterior probability that each galaxy from a catalog within a defined angular search cone is the transient's host.
\ac{SPLASH} associates \ac{SNe} with hosts in either \ac{GLADE} \citep{GLADE_2018} or \ac{DECaLS} \citep{Dey_2019} and then queries \pstarr for photometry using a cone search, as it is the same dataset used to train \ac{SPLASH}'s host property inference (see Section~\ref{subsec:host_prop_NN}).
If \texttt{Pröst} cannot find a host galaxy or claims that a \ac{SN} is hostless, \ac{SPLASH} does not classify it.

\texttt{Pröst}'s posterior probability estimate can be conditioned on any combination of the redshift of the transient, its fractional radial offset from a galaxy (the angular offset relative to the directional light radius of the galaxy in the direction of the transient; see \citet{gupta_2016} for details), and the galaxy's intrinsic brightness.
For each of these quantities, the user supplies a prior distribution. For fractional offset and host brightness, the user also defines a likelihood based on the transient being associated and the survey from which the transient was detected (informed by archival SN samples). The redshift likelihood is assumed Gaussian and calculated empirically by comparing each galaxy's redshift and its reported uncertainties to the redshift of the queried transient. Monte-Carlo samples of the association are drawn from the uncertainties in each measured property, and the galaxy associated in the largest number of \textit{N} trials is chosen as the host.
If a transient's redshift is not provided, \texttt{Pröst} will marginalize over a given prior.
For \ac{SNe} where redshifts are not given by the user, \ac{SPLASH} takes the redshift value from \texttt{Pröst}.

We choose relatively uninformed priors and physically motivated likelihood functions.
For the fractional offset and absolute magnitude, we assume uniform priors over $[0, 10]$ and $[-30, -10]$, respectively.
For the redshift prior, we select a half-normal distribution with mean $10^{-4}$ and variance $0.5$ to reflect low redshift selection preferences.
We use the likelihood function of the gamma distribution with a parameter of $0.75$ for the fractional offset to reflect how \ac{SNe} observed near the centers of galaxies are most likely to be hosted by those galaxies, as is the default in \texttt{Pröst}.
We adopt the absolute magnitude likelihood function from \citet{Li_2011} where the \ac{SN} rate scales as $~0.1\rm{L}_{\rm{host}}$ and $\rm{L}_{\rm{host}}$ is the estimated host luminosity in units of $10^{10} \ \rm{L}_\odot$.
$\rm{L}_{\rm{host}}$ is estimated by converting the catalog's recorded galaxy magnitude to absolute magnitude using the given redshift estimate.

\subsubsection{Probability of Chance Coincidence}
\label{subsubsec:Pcc}

For testing purposes, we create validation datasets by associating \ac{OACAPI} \ac{SNe} with host galaxies from \citet{Zou_2021} and \ac{SNe} from \ac{BTS} \& \ac{YSE} with hosts from \pstarr.
To reduce the computational cost of compiling these validation sets, we perform host association using methodology from \citet{Bloom_2002} instead of \texttt{Pröst}.
Originally used for short gamma ray burst association, \citet{Bloom_2002} calculates the ``probability of chance coincidence" which is defined as the probability of a \ac{SN} being a given effective radius $\delta R$ from a galaxy's flux-weighted center and \textit{not} being its host.

To calculate the probability of chance coincidence, we begin by finding the number density of galaxies brighter than a magnitude $m$,
\begin{equation}
    n(\leq m) = \frac{1}{0.33\ln(10)}10^{0.33(m-24)-2.44}\ \rm{arcsec}^{-2}
    \label{eq:n_lt_m}
\end{equation}
which is based on deep optical results from \citet{Hogg_1997} and \citet{Beckwith_2006}. Then, the probability of chance coincidence follows from a two dimensional Poisson process:
\begin{equation}
    P_{\rm{cc}} \equiv P(\delta R) = 1 - \exp\left[-\pi (\delta R)^2 n(\leq m)\right]
\end{equation}
where $\delta R = (R_0 + 4R_{\rm{half}}^2)^{(1/2)}$ is a conservative estimate for the effective radius with $R_0$ as the radial separation of the \ac{SN} from the galaxy center and $R_{\rm{half}}$ as the galaxy's half light radius.
We choose $P_{\rm{cc}} < 0.1$ as the criterion for a galaxy-\ac{SN} pair, and we select the candidate with the minimum chance of coincidence if multiple galaxies have $P_{\rm{cc}} < 0.1$.
If the criterion of $0.1$ is not met by any galaxies, the given \ac{SN} is considered an orphan and is not included.


\begin{figure}
  \centering
  \includegraphics[width=\linewidth]{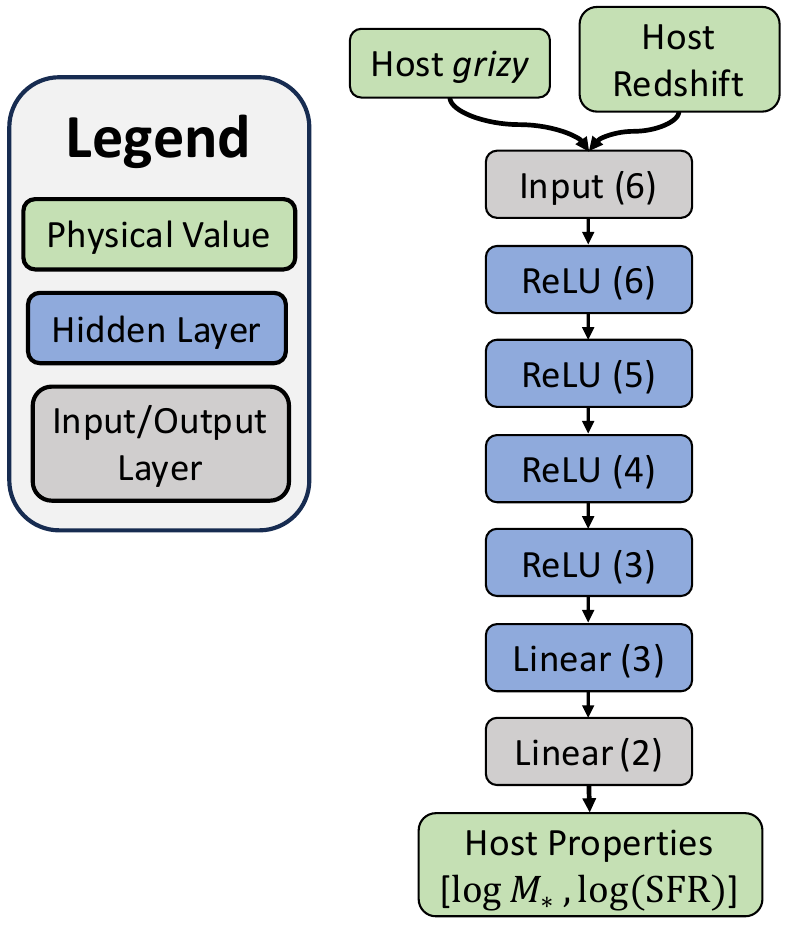}
  \caption{The architecture of the host property inference MLP.
  Each layer is depicted as a block with arrows representing the inputs and outputs.
  The number in parenthesis inside each block is the number of nodes in the layer, and all adjacent layers are fully connected.}
  \label{fig:nn_architecture}
\end{figure}

\subsection{Host-Property Inference with Deep Learning} \label{subsec:host_prop_NN}

We use a \ac{MLP} \ac{NN} to infer host galaxy properties from their $grizy$ absolute magnitudes and redshift.
\acp{MLP} are comprised of an input layer, several ``hidden" layers, and an output layer.
The interconnected layers of \acp{MLP} possess nonlinear ``activation" functions that allow them to model complex relationships, rendering them effective for classification, regression, and pattern recognition tasks.
They are very efficient to optimize because they are differentiable, which is a major reason why they have become a preferred tool for solving a wide range of computational problems in recent years.

To train the host property \ac{NN}, we use an $80$-$20$ train-test split of the \citet{Zou_2022} catalog.
Because \photozs can be highly uncertain, we select our training set from the $84,672$ galaxies for which \citet{Zou_2022} provides \specz measurements.
The input layer of our \ac{NN} uses galaxy magnitudes as input and conducts gradient descent on a loss function based on the \ac{MSE} of host property predictions, defined by
\begin{equation}
    \rm{Loss}(\mathbf{y}) = \frac{1}{N} \sum_{i=1}^N \frac{1}{\mathbf{\sigma}_i} (\mathbf{y}_i - \mathbf{\hat{y}}_i)^2
    \label{eq:hp_loss_fn}
\end{equation}
where $\rm{N}$ is the number of galaxies in a batch and $\mathbf{y}_i = [\log(\rm\ac{M_s})_i, \log(\rm{\ac{SFR}})_i]$ is a vector of the $i$th host galaxy's properties.
Assuming that the spectroscopic errors are negligible, we set the spectroscopic errors to an arbitrarily small value. 
Alternatively, per \cite{Zou_2022}, we set the uncertainty on \photozs to
\begin{equation}
    \sigma_z = \frac{1}{2} (z_{\rm{max}} - z_{\rm{min}})
    \label{eq:sigma_z}
\end{equation}
where $z_{\rm{max}}$ and $z_{\rm{min}}$ are given constraints on the redshift.

In Figure~\ref{fig:nn_architecture}, we display the architecture of our \ac{MLP}: a $6$-node input layer (corresponding to $g$, $r$, $i$, $z$, $y$, and redshift), followed by four hidden layers with \ac{ReLU} activation functions, a linear hidden layer, and a linear output layer with two nodes corresponding to \ac{M_s} and \ac{SFR}.
We use the decaying AdamW stochastic optimizer to train our network (see the \texttt{PyTorch} documentation and \citet{loshchilov_2019} for details).

We tune the \ac{MLP} with a brute-force grid search over batch size, number of epochs, nodes per layer, learning rate, and number of linear hidden layers at the end of the network.
Using this grid search, we select the architecture in Figure~\ref{fig:nn_architecture}(a), a learning rate of $0.01$, and batch size of $2048$ for $10,000$ training epochs.
We use early stopping with a $100$-epoch maximum for non-decreasing test loss.

\subsection{Random Forest Classification} \label{subsec:random_forest}
We train a \ac{RF} to infer the classes of \ac{SNe} given \ac{NN}-inferred host \ac{M_s} and \ac{SFR} values along with the redshift and host-\ac{SN} angular separation.
\acp{RF} are an ensemble learning architecture known for their robustness at handling complex, high-dimensional datasets.
By constructing a large number of decision trees and aggregating their outputs, they achieve high accuracy while mitigating overfitting.
\acp{RF} are favored among classification methods for their efficiency with large datasets and their interpretability.

We use a relatively simple \ac{RF} model from the \texttt{scikit-learn} package \citep{scikit-learn} comprised of $1000$ trees.
We adopt the standard Gini impurity as our loss function.

Throughout this paper, we will use stratified $k$-fold cross-validation to evaluate the performance of our pipeline.
This method splits data into $k$ sets (folds) while ensuring that each split has a similar class distribution.
It then iterates through the folds, using one fold to test and the rest to train.
This process is repeated $k$ times, allowing each fold to be used as a test set once and providing a comprehensive and unbiased assessment of \ac{RF} performance across samples.
We will sometimes opt for regular $k$-fold cross-validation, which is identical to the stratified variant except that it skips the requirement that each fold has a similar distribution of classes.

\subsection{Evaluation Metrics} \label{subsec:evaluation_metrics}
We use several different metrics to quantify inference and classification performance, which we define below.

The error and fractional error of inferences are
\begin{equation}
    \rm{Error} = |\hat{y} - y|
    \label{eq:err}
\end{equation}
\begin{equation}
    \rm{Fractional \ Error} = \left| \frac{\hat{y} - y}{\rm{max}(\hat{y},\ y)} \right|
    \label{eq:frac_err}
\end{equation}
where $y$ is the true value and $\hat{y}$ is the value inferred by our MLP.
Note that we use the maximum of the inferred and observed values in the denominator of Equation~\ref{eq:frac_err} to account for near-zero floating point errors that cause the expression to blow up.

The \ac{RMSE} of a set of inferences is
\begin{equation}
    \rm{RMSE} = \sqrt{\sum_{i=1}^N \frac{\rm{Error}^2}{N}}
    \label{eq:RMSE}
\end{equation}
where $N$ is the size of our test set and error is defined in Equation~\ref{eq:err}.
We use \ac{RMSE} to compare the relative uncertainty of inferences to the uncertainty of true values---specifically, we compare this value to the measured uncertainty divided by the measured value to get a sense of how well our tuned model is given the quality of the data.

We calculate the mean purity in the $k$-fold stratified cross-validation to measure the classification performance, which we describe in \ref{subsec:random_forest}.
Purity and completeness---which are used interchangeably with precision and recall---are defined by
\begin{equation} \label{eq:purity}
    \rm{Purity} = \frac{\rm{TP}}{\rm{TP} + \rm{FP}}
\end{equation}
\begin{equation} \label{eq:completeness}
    \rm{Completeness} = \frac{\rm{TP}}{\rm{TP}+\rm{FN}}
\end{equation}
where TP is the number of true positives, FP is the number of false positives, and FN is the number of false negatives for a set of class inferences. Additionally, accuracy is the number of correct predictions divided by the total number of predictions:
\begin{equation}
    \rm{Accuracy} = \frac{TP + TN}{TP + TN + FP + FN}
\end{equation}
where TN is the number of true negatives.

Finally, to compare \ac{SPLASH}'s performance with other classification methods, we calculate the F1-score.
The F1-score is the harmonic mean of a classifier's purity and completeness,
\begin{equation} \label{eq:f1_score}
    \rm{F}1 = \frac{2}{\rm{completeness}^{-1} + \rm{purity}^{-1}}
\end{equation}
where purity and completeness are defined in Equation~\ref{eq:purity} and Equation~\ref{eq:completeness}, respectively.
F$1$-scores can range from $0.0$ as the worst to $1.0$ being perfect classification performance.
We report the class-averaged and ``weighted" F1-scores, in which each class's F1-score is given a weight proportional to the number of \ac{SNe} of that class.

\section{Performance} \label{sec:pipeline_performance}


\subsection{Host Property Inference} \label{subsec:host_prop_NN_performance}
\begin{figure}
  \centering
  \includegraphics[width=\linewidth]{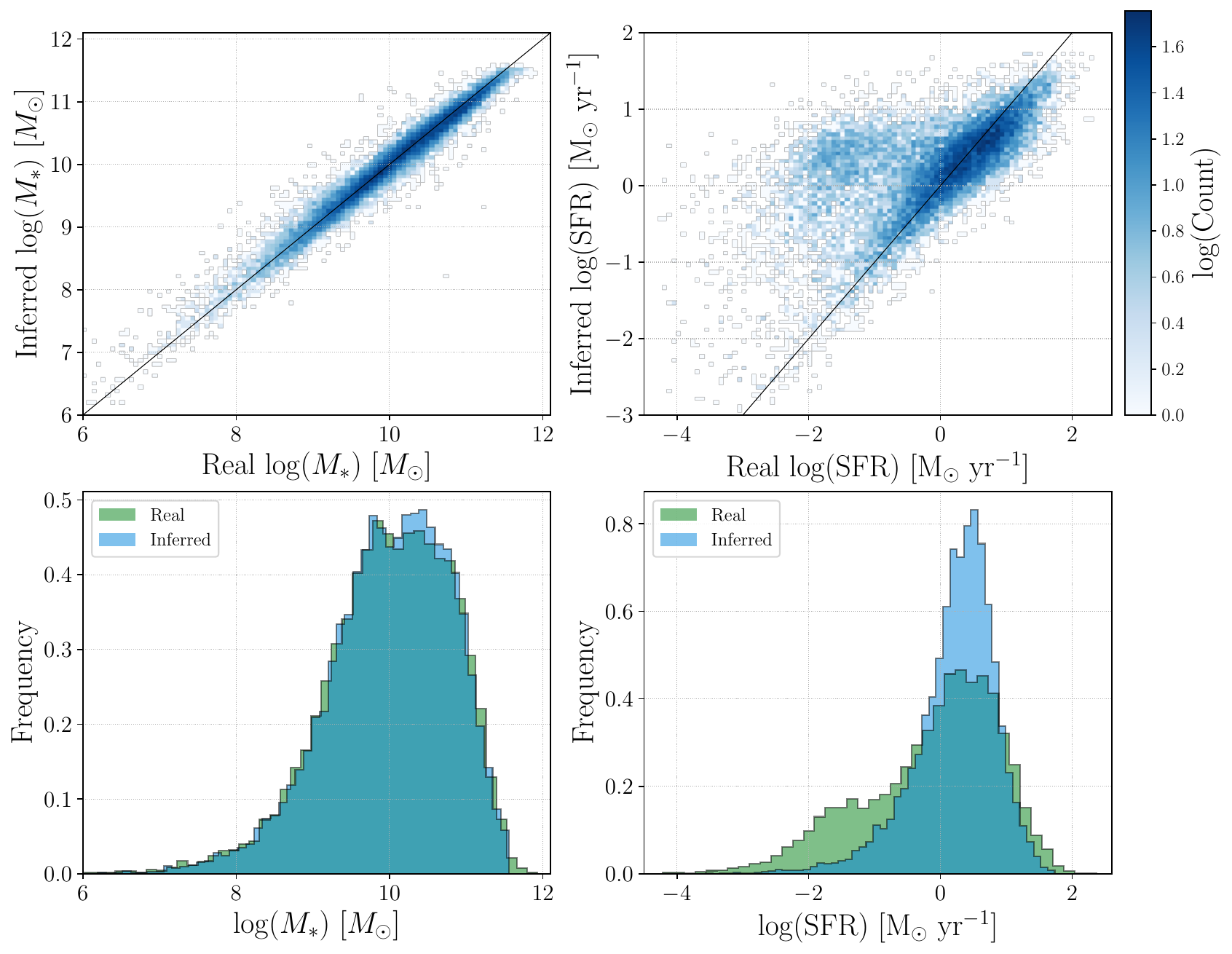}
  \caption{The real versus predicted values of the stellar mass and star formation rate for the $16,935$-galaxies in the test set with \speczs.
  \textbf{Top row:} A heatmap of the real versus predicted distribution of the host galaxy properties colored by log count.
  We include a black line with slope $1$ that goes through the origin to represent where inferences should fall if they were exactly accurate.
  \textbf{Bottom row:} The normalized histograms of the real and inferred distributions.}
  \label{fig:real_v_preds}
\end{figure}

\ac{SPLASH} performs well at inferring \ac{M_s} and \ac{SFR} from galaxy $grizy$ photometry and redshift.
In Figure~\ref{fig:real_v_preds}, we show the true versus NN-inferred distributions of \ac{M_s} and \ac{SFR} of the \citet{Zou_2022} host galaxy test set, and we see broad agreement between true and our photometrically inferred values.
The metrics listed in Table~\ref{tab:metrics} confirm that \ac{SPLASH} accurately infers galaxy properties.
We see a median fractional error (defined in Eq.~\ref{eq:frac_err}) of $\simeq0.01$ and $\simeq0.2$ for \logM and \logSFR, respectively.

For a small population of high-\ac{M_s} galaxies, \ac{SPLASH} overpredicts \ac{SFR}.
Overpredictions of the \ac{SFR} can be observed as a blue cluster and low-\ac{SFR} bump in the upper right and lower right panels of Figure~\ref{fig:real_v_preds}, respectively.
This group of galaxies makes up roughly $15\%$ of the dataset, and corresponds to a region with $\log (\textrm{sSFR})$ less than $10^{-12}$ where $\rm{sSFR} = \frac{\rm{SFR}}{M_*}$ is the specific \ac{SFR}.
Galaxies with $\rm{sSFR} \leq 10^{-12}$ are effectively quiescent, making it challenging to record accurate measurements below this limit.
Beyond the fact that there are negligible physical differences between quiescent galaxies, \ac{SPLASH} is designed to find \ac{CCSNe} which are seldom found in quenched galaxies, so it is okay for \ac{SPLASH} to lose fidelity in this region.

\begin{table*}[t!]
\centering
\begin{tabular}{l|c|c|c|c|c|c}
\hline
\multirow{2}{*}{\textbf{Statistic}} & \multicolumn{2}{c|}{\textbf{Fractional Error}} & \multicolumn{2}{c|}{\textbf{Error}} & \multicolumn{2}{c}{\textbf{Measured Uncertainties}} \\ \cline{2-7} 
                                 & \makebox[\calcfactor{.15}][c]{\textbf{log($\mathbf{M_*}$)}}
                                 & \makebox[\calcfactor{.15}][c]{\textbf{log(SFR)}}
                                 & \makebox[\calcfactor{.15}][c]{\textbf{log($\mathbf{M_*}$)}}
                                 & \makebox[\calcfactor{.15}][c]{\textbf{log(SFR)}}
                                 & \makebox[\calcfactor{.15}][c]{\textbf{log($\mathbf{M_*}$)}}
                                 & \makebox[\calcfactor{.15}][c]{\textbf{log(SFR)}}
                                 \\ \hline \hline
Mean                             & 0.01       & 0.21       & 0.12       & 0.56       & 0.14       & 0.30      \\ 
Median                           & 0.01       & 0.11       & 0.08       & 0.29       & 0.14       & 0.16      \\
Stdev                            & 0.01       & 0.25       & 0.12       & 0.67       & 0.06       & 0.38      \\ \hline
\multicolumn{7}{c}{\textbf{Other Metrics}} \\ \hline

\multicolumn{7}{c}{%
  \makebox[\calcfactor{.33}][l]{\textbf{Metric}}%
  \hspace{\tabcolsep}\vline\hspace{\tabcolsep}%
  \makebox[\calcfactor{.33}][c]{\textbf{log($\mathbf{M_*}$)}}%
  \hspace{\tabcolsep}\vline\hspace{\tabcolsep}%
  \makebox[\calcfactor{.33}][c]{\textbf{log(SFR)}}%
}\\ \hline \hline

\multicolumn{7}{c}{%
  \makebox[\calcfactor{.33}][l]{Prediction RMSE}%
  \hspace{\tabcolsep}\vline\hspace{\tabcolsep}%
  \makebox[\calcfactor{.33}][c]{0.1682}%
  \hspace{\tabcolsep}\vline\hspace{\tabcolsep}%
  \makebox[\calcfactor{.33}][c]{0.8675}%
}\\
\multicolumn{7}{c}{%
  \makebox[\calcfactor{.33}][l]{Measured (Uncertainty)/(Value)}%
  \hspace{\tabcolsep}\vline\hspace{\tabcolsep}%
  \makebox[\calcfactor{.33}][c]{0.0121}%
  \hspace{\tabcolsep}\vline\hspace{\tabcolsep}%
  \makebox[\calcfactor{.33}][c]{1.4655}%
}\\ \hline

\end{tabular}
\caption{Summary metrics of the neural network performance. We show the error (Eq.~\ref{eq:err}), fractional error (Eq.~\ref{eq:frac_err}), and RMSE (Eq.~\ref{eq:RMSE}) for the neural network inferences of galaxy stellar mass and star formation rate. Note that Eq.~\ref{eq:frac_err} is a variation of fractional error used to account for floating point issues. We include measured uncertainties and measured uncertainty divided by true value magnitude for reference.}
\label{tab:metrics}
\end{table*}


\subsection{Classification} \label{subsec:rf_performance}

\begin{table*}[t]
\centering
\begin{tabular}{l|c|c|c|c|c|c|c}
\hline \multirow{2}{*}{\textbf{Dataset}} & \multicolumn{2}{c|}{\textbf{Binary F1}} & \multicolumn{2}{c|}{\textbf{3-Class F1}} & \multicolumn{2}{c|}{\textbf{5-Class F1}} & \multirow{2}{*}{\textbf{Accuracy}} \\ \cline{2-7}
                 & \textbf{Unweighted} & \textbf{Weighted} & \textbf{Unweighted} & \textbf{Weighted} & \textbf{Unweighted} & \textbf{Weighted} & \\
\hline \hline
Pan-STARRS       & $0.59 \pm 0.03$   & $0.72 \pm 0.02$ & $0.39 \pm 0.03$   & $0.68 \pm 0.02$ & $0.25 \pm 0.03$   & $0.67 \pm 0.02$ & $0.76 \pm 0.02$ \\
BTS \& YSE       & $0.69$             & $0.79$           & $0.35$             & $0.71$           & $0.30$             & $0.73$           & $0.76$ \\
Zou+2022         & $0.49 \pm 0.13$   & $0.72 \pm 0.08$ &                     &                   &                     &                   & $0.76 \pm 0.07$ \\
\end{tabular}
\caption{The binary F1-score, three-class F1-score, five-class F1-score, and accuracy achieved by \texttt{SPLASH} classification on the Pan-STARRS, BTS \& YSE, and \citet{Zou_2022} datasets.
Unweighted F1-scores and F1-scores weighted by the support are included.
Uncertainties are included where metrics are calculated by taking a mean across cross-validation trials, and some entries are left empty where no score was calculated.}
\label{tab:classification_results}
\end{table*}

\begin{figure}
  \centering
  \includegraphics[width=\linewidth]{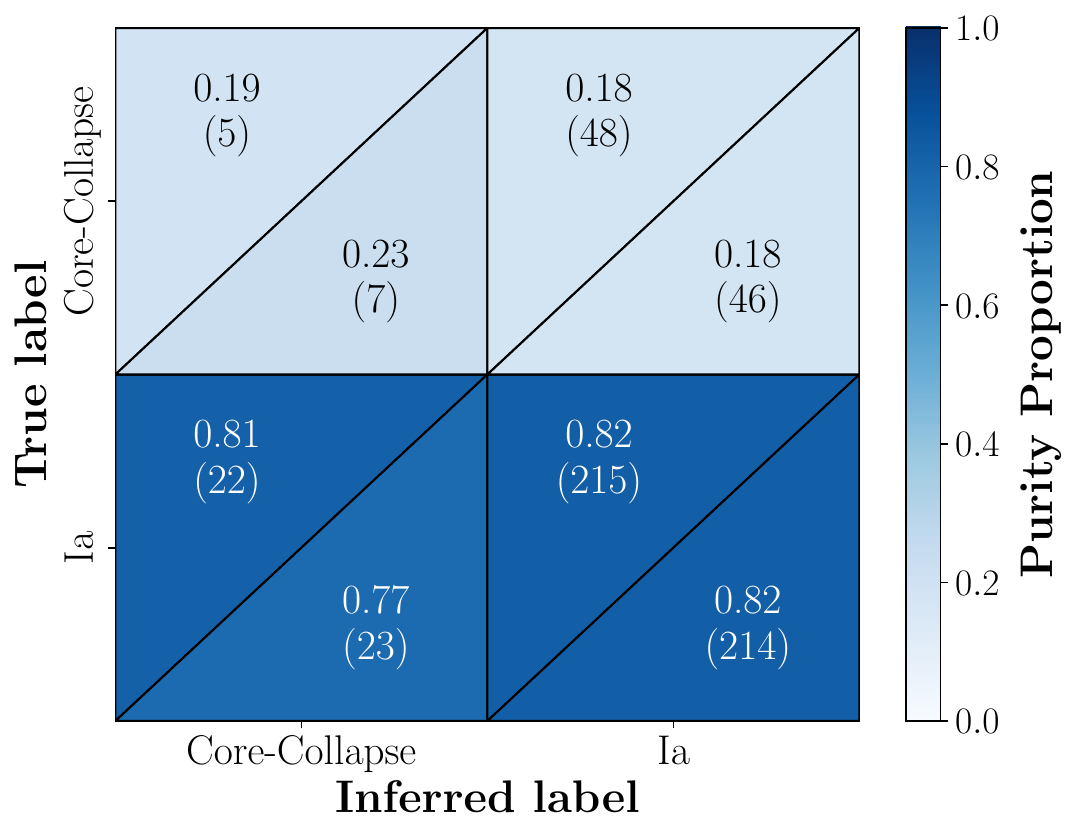}
  \caption{The cumulative type Ia vs. \ac{CCSN} confusion matrix across 20-fold stratified cross-validation of SNe with hosts in the \citet{Zou_2022} catalog.
  Each square in the grid shows the mean purity across folds, and the top and bottom triangles within each square correspond to the performance using NN-inferred and derived properties, respectively.
  The numbers of SNe in each category are included in parenthesis.}
  \label{fig:nn_inferred_v_true_cm}
\end{figure}

We show the cumulative binary confusion matrix for Type Ia \ac{SNe} and \ac{CCSNe} using NN-inferred and true (derived) properties across stratified $50$-fold cross-validation in Figure~\ref{fig:nn_inferred_v_true_cm}.
Classification with NN-inferred and derived properties are very similar, i.e., the purities between triangles in each region of the matrix differ by $\Delta \rm{Purity} \leq 0.04$.
Achieving such close performance with \ac{NN}-inferred properties serves as proof-of-concept for the upstream portion of \ac{SPLASH} that infers host properties because it indicates that the inferred properties capture the same information as the true values.

\begin{figure}
  \centering
  \includegraphics[width=\linewidth]{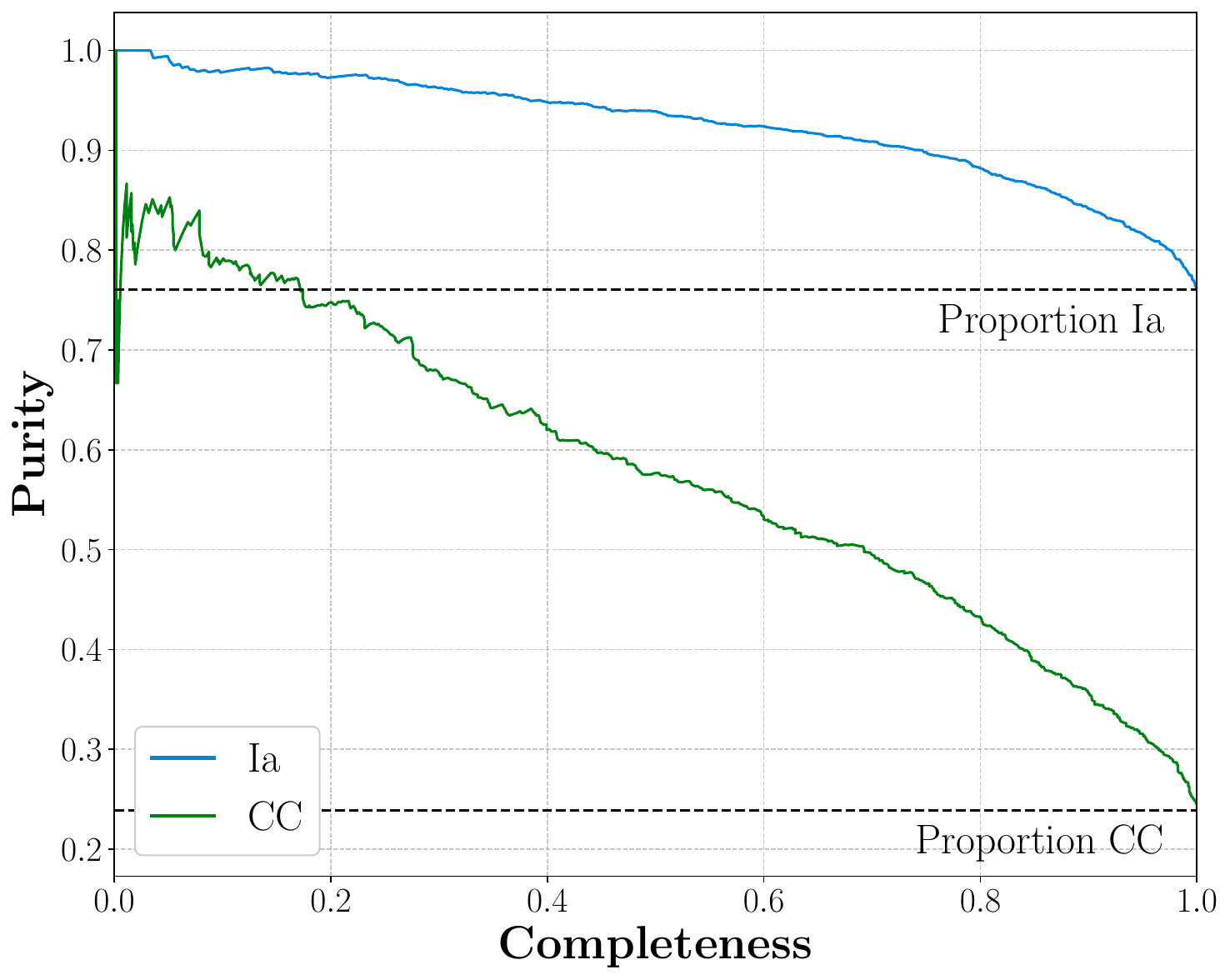}
  \caption{Purity versus completeness for \texttt{SPLASH} binary classification of \ac{SNe} from the BTS and YSE catalogs.}
  \label{fig:bts_yse_purity_v_completeness}
\end{figure}

As can be seen in the purity versus completeness plot in Figure~\ref{fig:bts_yse_purity_v_completeness}, adjusting sample completeness can dramatically increase \ac{SPLASH}'s classification purity.
In fact, \ac{SPLASH} achieved a purity of $\sim 80\%$ and $\sim 97.5\%$ for \ac{CCSNe} and Type Ia SNe, respectively, by limiting the sample to only the most confident $10\%$.
In Appendix Figure~\ref{fig:all_class_cm_by_thresh} we show the five-class confusion matrices for different completeness thresholds, and we observe that the purities of all classes are improved by lowering completeness.

To test whether the intermediate step of inferring galaxy properties makes \ac{SPLASH} a better classifier than if it just used photometry, we compare the results of our pipeline to a \ac{RF} that classifies \ac{SNe} from only the host \ac{SED}.
The photometry-based \ac{RF} performs slightly worse than \ac{SPLASH}, giving a class-weighted and unweighted F1-score of $71 \pm 1$\% and $59 \pm 2\%$, respectively, while \ac{SPLASH} achieves F1-scores of $75 \pm 1$\% and $63 \pm 2\%$ (see Figure~\ref{fig:nn_inferred_v_true_cm} for \ac{SPLASH}'s binary confusion matrix).
Thus, inferring galaxy properties as an intermediary step grounds predictions in intuitive, physical terms, but also gives a slight boost to classification performance.

\begin{figure*}
  \centering
  \includegraphics[width=\linewidth]{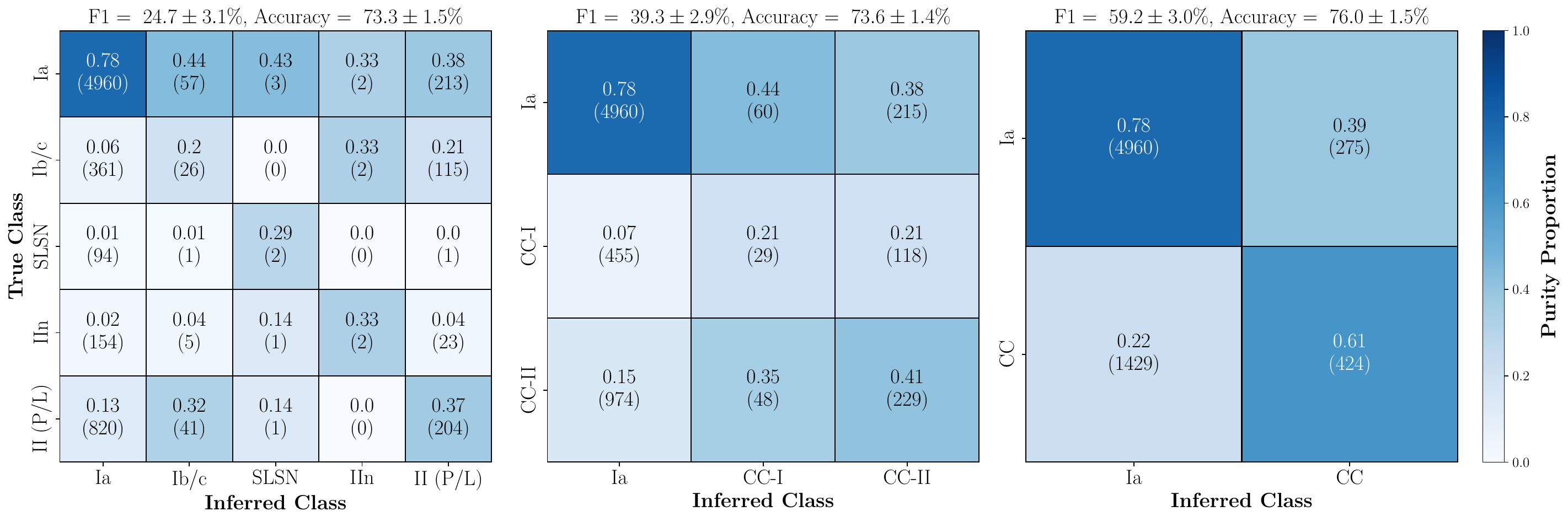}
  \caption{
  \textbf{Left:} The \ac{SPLASH} five-class (left), three-class (center), and two-class (right) confusion matrices using $20$-fold cross validation on the supernova dataset.
  The means and standard deviations across the folds of the unweighted F1-scores and accuracies are include above the matrices.
  The numbers of SNe in each group are included in parenthesis.
  }
  \label{fig:all_class_cm}
\end{figure*}

In Figure~\ref{fig:all_class_cm}, we show the cumulative five-class, three-class, and binary confusion matrices for \ac{SPLASH} across $20$-fold cross-validation for \ac{SNe} in the \pstarr catalog.
We list the class-weighted F1-score, unweighted F1-score, and accuracy for the five, three, and binary class breakdown in Table~\ref{tab:classification_results} for \pstarr as well as the \citet{Zou_2022} and BTS \& YSE datasets.
In the five-way task, \ac{SPLASH} achieves fair purities of $78\%$, $20\%$, $33\%$, $37\%$, and $29\%$ for Type Ia, I(b/c), IIn, II(P/L), and SLSNe, respectively.
\ac{SPLASH} achieves fair purities in the three-way task as well, yielding $78\%$, $21\%$, and $41\%$ for Type Ia, Type I \ac{CCSNe}, and Type II \ac{CCSNe}, respectively.
In the five-way task, the majority of misclassifications are between subclasses of \ac{CCSNe}, which is consistent with the fact that CCSNe occur in star-forming galaxies \citep{Li_2011, Hakobyan_2012, Childress_2013, Villar_2025}.
\ac{SPLASH} confuses \ac{SLSNe} with other subclasses of \ac{CCSNe} the least, which is probably because they prefer high-redshift, dwarf galaxies---the dimmest hosts that we expect to have in our catalog of \ac{SNe}.
For three-class classification, \ac{SPLASH} performs fairly well but still struggles to differentiate between the Type I and Type II subclasses of \ac{CCSNe}, producing modest purities of $21\%$ and $41\%$, respectively.
In binary classification we perform quite well, achieving a relatively high purity of $61\%$ for \ac{CCSNe} and $78\%$ for Type Ia.
Classification error among all classes comes from considerable overlap between classes in the host galaxy parameter space (see Figure~\ref{fig:pairplot}(b)), making the classification task inherently difficult, especially for minority classes.
Performance limitations in host-based classification will be discussed further in Section~\ref{sec:discussion}.

\section{Discussion} \label{sec:discussion}
\subsection{Classification Performance} \label{subsec:classification_performance}

\ac{SPLASH} demonstrates cutting-edge photometric \ac{SN} classification performance using only host galaxy photometry.
In the five-class classification task, \ac{SPLASH} achieves a F1-score of $30 \%$ on the BTS \& YSE dataset, somewhat less than the host-only F1-score of $36 \%$ from \citet{Villar_2025}.
For three-class classification, our F1-score of $35\%$ is below the $49\%$ from \citet{Villar_2025} and $48\%$ from \citet{gagliano_2023}.
The performance gap may stem from methodological differences: \citet{Villar_2025} uses a hierarchical classification model and \citet{gagliano_2023} leverages early \ac{SN} light curve measurements.
Our binary accuracy of $76\%$ exceeds the $68\%$ from \citet{astro_ghost} and $66\%$ from \citet{Villar_2025}, and our unweighted F1-score of $69\%$ is higher than the $66\%$ reported by \citet{Villar_2025}.
Notably, the datasets used to calculate these metrics are different for each study, so their direct comparison is not completely representative of the methods' relative performances.
In Table~\ref{tab:classification_results}, we list \texttt{SPLASH}'s test and validation performance metrics for all datasets used in this study (each of which is described in Section~\ref{subsec:datasets}).

Recent host-information-only \ac{SN} classification methods like \ac{SPLASH} all achieve relatively similar performance (e.g. \citet{astro_ghost, Qin_2024, Villar_2025}).
Using different approaches and architectures for host-based photometric classification has not appeared to substantially improve classification, implying that the field may have hit a performance limit that is intrinsic to the task of host-based classification itself.
In our case, Figure~\ref{fig:pairplot}(b) demonstrates that the properties of \ac{SN} hosts are not perfectly distinct in feature space, but are rather highly overlapped.
Although the properties of hosts are clearly correlated with the types of \ac{SNe} that they produce, this relationship is not absolute.
Future photometric classification methods would likely benefit from supplementing photometric information with other information about \ac{SNe} and their hosts to make substantial performance improvements.
For example, \citet{Baldeschi_2020} found that host morphology can be used to boost the purity of \ac{CCSN} classifications, and \citet{gagliano_2023} showed how early light curve information can improve performance for the three-way classification task.
It may also be interesting for future studies to explore how host properties in the \textit{local} region of a \ac{SN} impact classification results.

Our results show that \ac{SPLASH} is capable of obtaining pure samples of all five \ac{SN} classes that we consider.
In Figure~\ref{fig:bts_yse_purity_v_completeness}, we observe a purity of $\sim80\%$ for \ac{CCSNe} and $97.5\%$ for Type Ia when taking the most confident $10\%$ of our classifications in the BTS \& YSE dataset.
In Figure~\ref{fig:all_class_cm_by_thresh}, we show that as we increase the confidence threshold for the five-way classification task, the sample purity for all five classes increases except for \ac{SLSNe}, for which all examples are inferred to be Type Ia.
\ac{SPLASH} misclassifies \ac{SLSNe} as Type Ia due to the overwhelming dominance of Type Ia events in the dataset and the fact that \ac{SLSNe} comprise only $\sim 1\%$ of the sample.
Just increasing the classification threshold to have a minimum probability of $0.6$ yields a purity of $40\%$ for Type Ib/c, $44\%$ for Type II (P/L), and $100\%$ for Type IIn.
The random forest's feature importances of $0.24$, $0.22$, $0.22$, $0.31$ for angular separation, \ac{M_s}, \ac{SFR}, and redshift, respectively, are relatively similar, suggesting that all four metrics give approximately the same amount of insight into the nature of events hosted by galaxies.

Although we sacrifice completeness when requiring more confident classifications, the unprecedented number of \ac{SN} alerts produced by \ac{LSST} will be enough to collect large, pure samples of all classes of \ac{SNe}.
For example, based on \ac{LSST}'s projection of detecting one million \ac{SNe} per year, we estimate that \ac{SPLASH} can produce an $80\%$ pure sample of roughly $70$ \ac{CCSNe} per night using a completeness of $10\%$.
Acquiring such a large sample of targets \textit{every night} would already go beyond what can be followed spectroscopically, which underscores the importance of purity in our study, as opposed to completeness, in the era of Rubin.



\subsection{Selecting for Supernova Demographics With \ac{SPLASH}} \label{subsec:sn_demographics_splash}

Beyond classification, \ac{SPLASH} is valuable for \ac{SN} population studies as it allows users to select for desired host demographics.
Specifically, our method returns an inferred set of host \ac{M_s}, \ac{SFR}, and redshift, making it easy to filter datasets for subpopulations with specific host features.
For example, one might pose the question: does the distribution of \ac{SN} classes change as a function of \ac{M_s}?
Using \ac{SPLASH}, we infer that the proportion of \ac{CCSNe} with $\log(\rm{M}_*) < 8.5$ in the BTS \& YSE sample is $35\%$, and the fraction falls to $16\%$ for the population above the low-mass cut.
\ac{SPLASH}'s results in this case are relatively close to the dataset's true low-mass to high-mass change of $40\%$ to $23\%$, and \ac{SPLASH} helps lend insight into how this dataset demonstrates that \ac{CCSNe} prefer low-mass galaxies.

With a rate of $\sim500$ classifications per second on a modern laptop, \ac{SPLASH}'s speed gives it the capacity to tackle large datasets.
Rapid tools like \ac{SPLASH} will be crucial for conducting large population studies with the unprecedented volume of detections that we will see once LSST goes online.
Notably, \ac{SPLASH} will automatically perform host association using \texttt{Pröst} if a host galaxy is not provided.
Although automatic host association is extremely useful, it does add a few seconds to each of \ac{SPLASH}'s \ac{SN} classifications.
Future work may be interested in photometric classification based on transient cutout images, thereby eliminating the need for host association.

\section{Conclusions} \label{sec:conclusions}

In this paper, we introduced \ac{SPLASH}, a host-based \ac{SN} classification pipeline that rapidly classifies \ac{SNe} from host galaxy photometry.
By inferring stellar mass and star formation rate with a neural network and classifying \ac{SNe} with a random forest trained on these properties combined with host-transient angular separation and redshift, \ac{SPLASH} provides an interpretable, efficient, and scalable classification tool for the era of wide-field time-domain surveys.
Because \ac{SPLASH} relies solely on host information, it is particularly well-suited for very early-time classification when light curves are sparse or unavailable.
\ac{SPLASH} is actively classifying daily \ac{SN} alerts from the Transient Name Server on a publicly available website\footnote{\url{astrotimelab.com/_pages/splash.html}}.

SPLASH achieves performance comparable to other state-of-the-art host-based classifiers, with a binary (Type Ia vs. core-collapse) F1-score as high as $69\%$ and accuracy of $76\%$.
Furthermore, \ac{SPLASH} can produce highly pure samples of all \ac{SN} classes by requiring higher confidence classifications (i.e. lowering completeness), which is particularly important for selecting targets for spectroscopic or multi-wavelength follow-up.

Because \ac{SPLASH} relies solely on host photometry, it can classify transients immediately after detection when the transient is still young, even in the absence of \ac{SN} photometry.
Moreover, by accurately and rapidly inferring the properties of host galaxies ($\sim 500 \ \rm{galaxies} / \rm{sec}$) as an intermediate step for classification, \ac{SPLASH} can be used to select galaxies and \ac{SNe} for large population studies across distributions of physically-meaningful parameters.

Given the overwhelming rate of \ac{SN} detections expected from \ac{LSST}, scalable host-based methods like \ac{SPLASH} will be essential for early classification, prioritizing follow-up, and enabling large-scale population studies.
Future work may involve extending \ac{SPLASH} to include additional host properties like morphology or even latent representations from e.g., foundation models like AstroCLIP \citep{Parker_2024}, localized environmental metrics, or limited light curve information to further boost classification performance while maintaining efficiency and interpretability.

\begin{acknowledgments}
A.P.B. thanks Xingang Chen for running Harvard's Astronomy 98 course---in which a large portion of this work was developed---and for his guidance throughout the process.
The Villar Astro Time Lab acknowledges support through the David and Lucile Packard Foundation, National Science Foundation under AST-2433718, AST-2407922 and AST-2406110, as well as an Aramont Fellowship for Emerging Science Research.
This work is supported by the National Science Foundation under Cooperative Agreement PHY-2019786 (The NSF AI Institute for Artificial Intelligence and Fundamental Interactions, http://iaifi.org/). A.G.\ also acknowledges support from the AI Institutes Virtual Organization (AIVO) and the MIT Libraries.
The computations presented in this work were performed on the FASRC Cannon cluster supported by the FAS Division of Science Research Computing Group at Harvard University. 
\end{acknowledgments}

\software{Python 3.11.5 \citep{python}, NumPy \citep{Numpy}, Matplotlib \citep{Matplotlib}, PyTorch \citep{pytorch}, Astropy \citep{2013A&A...558A..33A,2018AJ....156..123A},  Scikit-learn \citep{scikit-learn}, Astro GHOST \citep{astro_ghost}, Pooch \citep{pooch}, \href{https://pypi.org/project/mastcasjobs/}{mastcasjobs}}


\bibliography{main_paper}{}
\bibliographystyle{aasjournal}



\appendix
\section{Classification With Less Completeness}

\begin{figure*}
  \centering
  \includegraphics[width=0.75\linewidth]{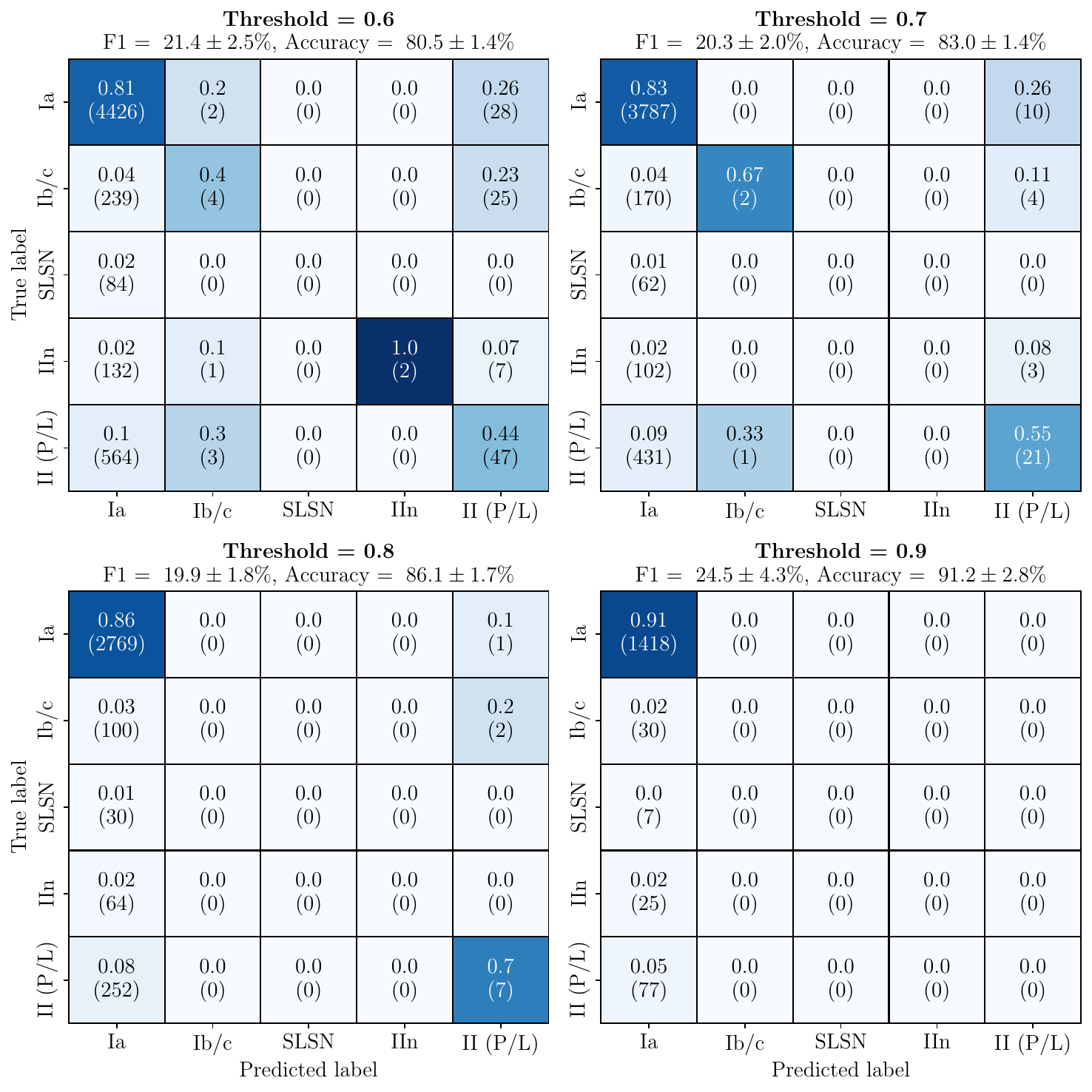}
  \caption{
  The \ac{SPLASH} cumulative confusion matrix with minimum classification probability thresholds of $0.6$, $0.7$, $0.8$, and $0.9$ across stratified 20-fold cross validation on the supernova dataset. The mean and standard deviation of the F1-score and accuracy across folds are shown above each matrix.
  }
  \label{fig:all_class_cm_by_thresh}
\end{figure*}

In Figure~\ref{fig:all_class_cm_by_thresh}, we show the \ac{SPLASH} 5-class confusion matrix for four different classification confidence requirements defined by minimum probability thresholds of $0.6$, $0.7$, $0.8$, and $0.9$.
We observe that as we require increasingly confident classifications, each \ac{SN} class's mean sample purity increases (except for \ac{SLSNe}, which we will discuss later in this section).
The purest samples that we obtain are $91\%$, $67\%$, $100\%$, and $70\%$ for Type Ia, Ib/c, IIn, and IIP/L, respectively.
The mean accuracy increases from $80.5 \pm 1.4\%$ with a threshold of $0.6$ to $91.2 \pm 2.8\%$ for a threshold of $0.9$ while the F1-score remains within the range of uncertainty for all four threshold values.

The sample completeness of our inferences falls as a result of requiring more confident classifications, but the volume of \ac{SN} alerts in the Rubin era will be so large that sample completeness is much less important than purity due to the constraints of spectroscopic follow-up resources (see Section~\ref{subsec:classification_performance}).
Unfortunately, using a minimum probability threshold $\geq 0.6$ yields no \ac{SLSNe} inferences at all, which is a result of the fact that \ac{SLSNe} make up only $\sim 1\%$ of the dataset (the smallest proportion of the five classes by more than a factor of two).

\end{document}